\documentclass[twocolumn]{el-author}
\usepackage{cite}
\usepackage{amsmath,amssymb,amsfonts}
\usepackage{algorithmic}
\usepackage{graphicx}
\usepackage{textcomp}
\usepackage{xcolor}
\usepackage{mathrsfs,amsmath}

\newcommand{\ua}{\uparrow}
\newcommand{\nc}{\newcommand}
\nc{\da}{\downarrow} \nc{\hc}{\hat{c}} \nc{\hS}{\hat{S}}
\nc{\bra}{\langle} \nc{\ket}{\rangle} \nc{\eq}{equation (\ref}
\nc{\h}{\hat} \nc{\hT}{\h{T}}\nc{\be}{\begin{eqnarray}}
\nc{\ee}{\end{eqnarray}}\nc{\rd}{\textrm{d}}\nc{\e}{eqnarray}\nc{\hR}{\hat{R}}\nc{\Tr}{\mathrm{Tr}}
\nc{\tS}{\tilde{S}}\nc{\tr}{\mathrm{tr}}\nc{\8}{\infty}\nc{\lgs}{\bra\ua,\phi|}\nc{\rgs}{|\ua,\phi\ket}
\nc{\hU}{\hat{U}}\nc{\lfs}{\bra\phi|}\nc{\rfs}{|\phi\ket}\nc{\hZ}{\hat{Z}}\nc{\hd}{\hat{d}}\nc{\mD}{\mathcal{D}}
\nc{\bd}{\bar{d}}\nc{\bc}{\bar{c}}\nc{\mc}{\mathcal}\nc{\ea}{eqnarray}\nc{\mG}{\mathcal{G}}\nc{\bce}{\begin{center}}
\nc{\ece}{\end{center}}

\begin{document}

\title{Theoretical analysis of the non-uniform OAM source based on an equivalent source transformation}

\author{Zelin Zhu, Shilie Zheng, Xiaowen Xiong, Yuqi Chen, Xiaofeng Jin, 
Xianbin Yu and Xianmin Zhang}

\abstract{Orbital angular momentum (OAM) has gradually become a research hotspot in radar, communication and other fields because of its phase distribution. However it encounters some problems in the real practice owing to its inherent characteristics. Recently OAM mode group has been proposed to solve these problems. In this paper, a non-uniform OAM source is used to generate the OAM  mode-group (MG) directly. The radiation field is analyzed based on an equivalent source transformation. Taking the partial arc transmitting (PAT) scheme as an example, the non-uniform source is equivalenced as a group of uniform OAM sources by calculated its Fourier transform. This work provides theoretical guidance for the design of non-uniform OAM antenna in the RF domain.}

\maketitle
\section{Introduction}

Thanks to its special phase distribution, the orbital angular momentum (OAM)-carrying beam has been proved to have the application prospect in the fields of communication, radar imaging, and etc, in radio frequency \cite{Mohammadi2010Orbital}. The central dark zone and the severe beam divergence make it difficult to maintain the orthogonality in multiplexing when the receiving aperture is limited. So OAM mode-group (MG) which consists of organized multimode OAM-carrying beams has been proposed\cite{Zheng2018Realization}. OAM MG can maintain the vortex characteristics of OAM-carrying beam while increase the directivity. Moreover, different OAM MGs can have quasi-orthogonality among the main lobe. Thus, the sampling receiving can carried out in the main lobe instead of the whole aperture for the conventional OAM beams, which can simplify the receiving end greatly. 
 These advantages are beneficial to the OAM based multiplexing in wireless communication\cite{cl2020MGmimo}. 

 Most of the already published OAM MG generation schemes were realized by the superposition of different OAM waves generated by separate OAM antennas, which is a bit bulky and complex. Hence a compact MG antenna which can generate OAM MG directly is high anticipated. In optical domain, a restriction of the angular range within an optical beam profile can generate OAM sidebands on the transmitted light \cite{Jack2008Angular}. Based on the fact that OAM-carrying wave can be generated by traveling-wave circular loop antenna\cite{xiong2020oam} or circular array antenna\cite{ISI:000492854000062}, it is naturally to think if the OAM MG can be generated by the non-uniform circular loop antenna. Partial arc transmitting (PAT) scheme is one example of the non-uniform traveling-wave circular loop antenna whose partial arc is with no feeding while the remaining part keeps the original feeding. The circular antenna array whose array element is fed by diverse power, 0 is the extreme case equivalent of no array element there can also be regarded as the non-uniform traveling-wave circular loop antenna. 
 
 \begin{figure}[htp]
\centering
  \includegraphics[width=3.5in]{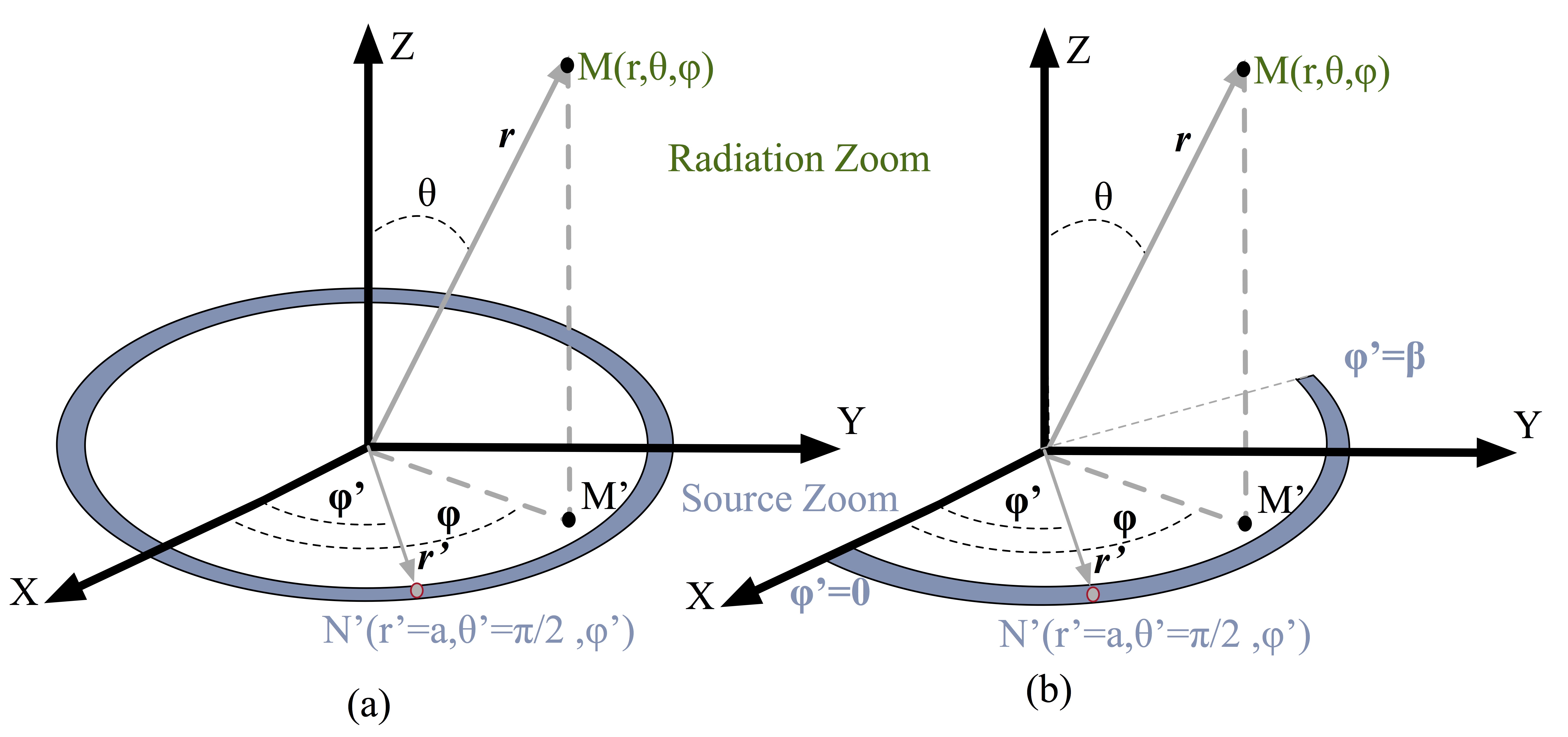}
  \caption{Different OAM source models: (a) the uniform OAM source (b) a kind of the non-uniform OAM source: partial arc transmitting}
  \label{p1}
\end{figure}

In this paper, the non-uniform traveling-wave circular loop antenna is proposed to generate the OAM MG. Taking the PAT scheme as an example, the farfield radiation of the non-uniform OAM source is theoretically analyzed. It is found that the radiation source can be equivalent to a sum of a series of uniform OAM sources, whose weights can be calculated by the Fourier transform (FT) of the amplitude distribution of the non-uniform source. The radiation field is thus the sum of the corresponding amplitude modulated OAM wave. Using this idea, the OAM MG can be effectively constructed. The main lobe of the OAM MG radiation and its OAM spectrum are also analyzed. This work not only provides a theoretical guidance for the design and optimization of the OAM MG antenna, but also helps the analysis of the non-uniform antenna in the RF domain.

\section{Analysis}
Fig.\ref{p1}(a) shows the uniform OAM source i.e. the traveling-wave circular loop antenna, $\vec I_{l}(l,\varphi')=I_{0}e^{-jl\varphi'}$ which can generate single mode $l$ OAM-carrying beam. The non-uniform OAM source $\vec I_{n}(l,\varphi')=I_{n}(\varphi')e^{-jl\varphi'}$ has a non-uniform amplitude distribution of $I_{n}(\varphi')$. Fig.\ref{p1}(b) shows the model of partial arc transmitting scheme which can be regarded as a uniform OAM source covered by a ring mask. 

When the uniform OAM source is fed, the farfield $\vec E_{l}(l,\theta,\varphi)$ at the point $M(r,\theta,\varphi)$ can be described as:
\begin{equation}\label{a1}
\begin{aligned}
  \vec E_{l}(l,\theta,\varphi)&=-\frac{\mu_{0}\omega}{4\pi}\int^{2\pi}_{0} \frac{I_{0}e^{-jl\varphi'}e^{jk|\vec r-\vec r'|}}{|\vec r-\vec r'|} d\varphi'\\
  &\approx C_{0}(-j)^{-l}J_{l}(kasin\theta)e^{-jl\varphi}\\
  &=C_{0}J_{|l|}(kasin\theta)e^{-jl(\varphi-\frac{l}{|l|}\frac{\pi}{2})}
  \end{aligned}
\end{equation}
where $C_{0}\propto\frac{\mu_{0}I_{0}\omega e^{jkr}}{4\pi r}$, $\mu_{0}$ is the permeability of free space, $\omega$ is the angular frequency and $k=\omega/c$ is the wave number. In the farfiled, $C_{0}$ is a constant term who is unrelated to $\theta$ and $\varphi$. According to the dual relation between the azimuthal domain and the mode domain\cite{JhaFourier}, the FT of the source $\vec I_{l}$ is
\begin{equation}
\label{a2}
  M\lbrack I_{l}(l)\rbrack=\mathscr{F}[\vec I_{l}(l,\varphi')]=\delta(l)
\end{equation}

When it comes to the PAT source, whose radiation model is shown in Fig.\ref{p2}(b), its source $\vec I_{n}(l,\varphi')$ is equal to $\vec I_{l}(l,\varphi')$ with certain mode $l_{0}$ multiplied by a gate function $g(\varphi')$:
\begin{equation}\label{a3}
  g(\varphi')=
  \begin{cases}
       1, & 0<\varphi'<\beta \\\\
        0, & \beta<\varphi'<2\pi
    \end{cases}
\end{equation}
where $\beta$ is the arc angle of $\vec I_{n}(l,\varphi')$. So the amplitude distribution function of the PAT source is $I_{n}(\varphi')=g(\varphi')I_{0}$. At this time, the radiation field $\vec E_{n}(\theta,\varphi)$ satisfies:
\begin{equation}
\label{a4}
     \vec E_{n}(\theta,\varphi)=-\frac{\mu_{0}\omega}{4\pi}\int^{2\pi}_{0} \frac{I_{n}(\varphi')e^{-jl_{0}\varphi'}e^{ik|\vec r-\vec r'|}}{|\vec r-\vec r'|} d\varphi'
\end{equation}

Compared with Eq.\ref{a2}, the FT of the source $\vec I_{n}(l,\varphi')$ can be calculated as
\begin{equation}
\label{a5}
\begin{split}
  M\lbrack I_{n}(l)\rbrack&=\mathscr{F}[g(\varphi')\vec I_{l}(l_{0},\varphi')]=Sinc(\frac{\beta l}{2})e^{jl\frac{\beta}{2}}*\delta(l_{0})
  \\&=Sinc(\frac{\beta (l-l_{0})}{2})e^{jl\frac{\beta}{2}}
  \end{split}
\end{equation}

Eq.\ref{a5} means that the non-uniform source $ \vec I_{n}$ is equivalent to a sum of a series of single mode source $\vec I_{l}(l)$ with weight function $M\lbrack I_{n}(l)\rbrack$:
\begin{equation}
\label{a6}
\begin{split}
 \vec I_{n}(l)= \sum_{l=-\infty}^{\infty} Sinc(\frac{\beta (l-l_{0})}{2})e^{jl\frac{\beta}{2}}\vec I_{l}(l,\varphi') 
 \end{split}
\end{equation}

Substituting Eq.\ref{a1} and Eq.\ref{a6} into Eq.\ref{a4}, the radiation field of the source $\vec I_{n}(l,\varphi')$ can be organized into:
\begin{equation}
\label{a7}
\begin{split}
 \vec E_{n}(\theta,\varphi)=C_{0}\sum_{l=-\infty}^{\infty} Sinc(\frac{\beta (l-l_{0})}{2})J_{l}(kasin\theta)e^{jl(\varphi-\varphi_{0})}
 \end{split}
\end{equation}
where $\varphi_{0}=\frac{\beta}{2}+\frac{l}{|l|}\frac{\pi}{2}$. The above equation shows that all modes have the same initial phase $\varphi_{0}$, and the main lobe in the $\varphi$ direction is $\varphi_{0}$ because of the in-phase stacking here. From the obtained radiation field, the main lobe direction $\theta_{0}$ in the $\theta$ direction can be found by calculating the maximum point. It's worth mentioning that $\varphi_{0}$ only depends on position of the partial arc and the amount of arc angle $\beta$.   
\begin{equation}
\label{a8}
\theta_{0}=\mathop{\arg\max_{ \theta}}{E_{n}(\theta,\varphi)}
\end{equation}

In the RF domain, the receiving antenna is usually placed at the maximum radiation ring, hence the OAM spectrum at the mainlobe direction $\theta=\theta_{0}$ meets:  
\begin{equation}
\label{a9}
|A_{l}|=|Sinc(\frac{\beta(l-l_{0})}{2})J_{l}(kasin\theta_{0})|
\end{equation}

According to Eq.\ref{a8} and Eq.\ref{a9}, the radiation characteristics of PAT, such as $\varphi_{0}$, $\theta_{0}$ and $|A_{l}|$ can be controlled by adjusting parameters $a$, $\beta$ and $l_{0}$.   
\section{Simulation}

\begin{figure}[htb]
 \centering
 \includegraphics[width=3.5in]{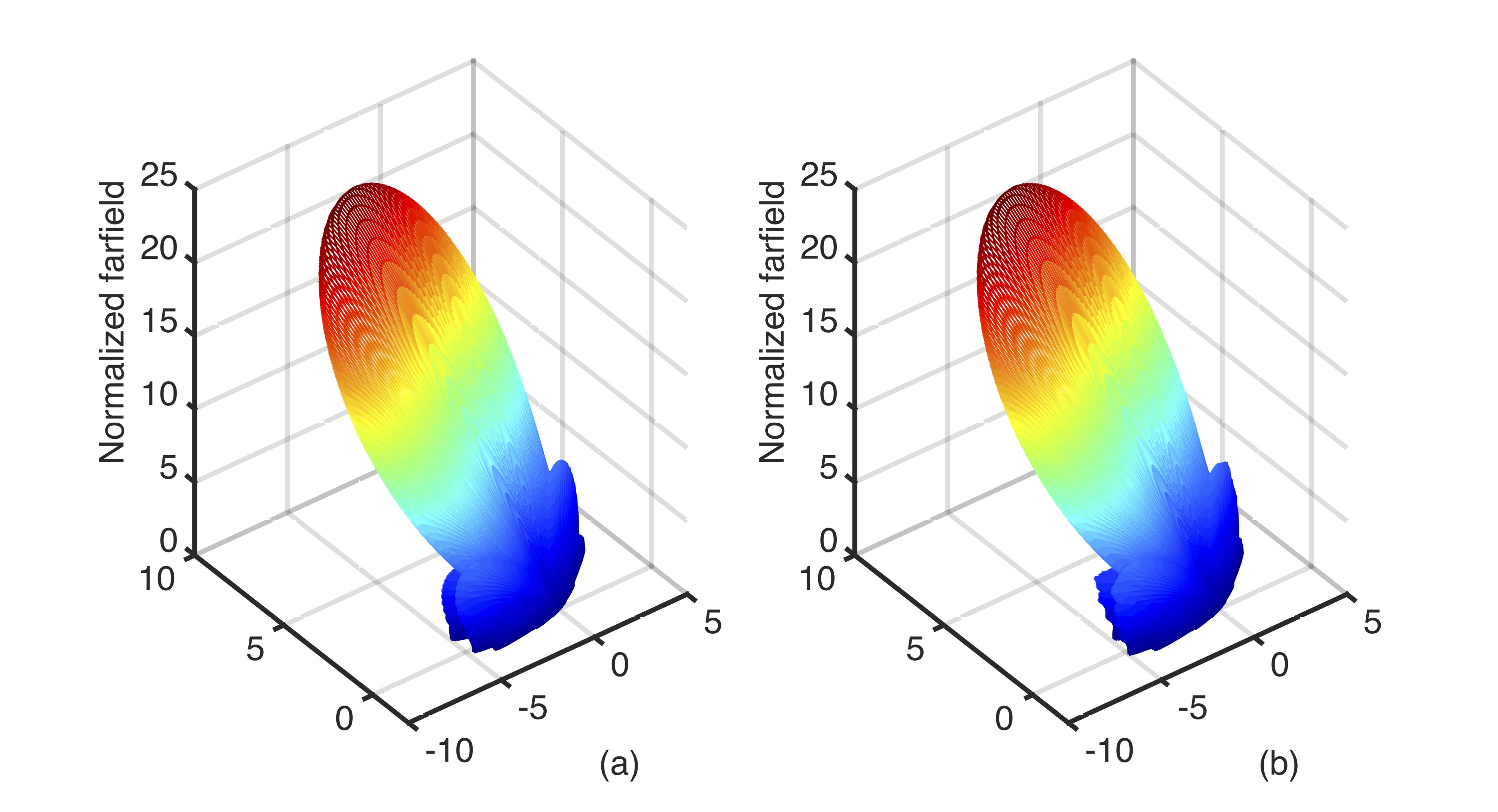}
 \caption{Numerical result of the farfield: (a) Calculated by Eq.\ref{a4} (b) Calculated by Eq.\ref{a7} }
 \label{p2}
\end{figure}

As an example, a one-nine aperture PAT antenna is designed, whose parameter list is shown in Tab.\ref{t1}\cite{xiong2020oam}. Fig.\ref{p2}(a) and (b) show the numerical results of the farfield of this antenna calculated by Eq.\ref{a4} and Eq.\ref{a7}, respectively. Obviously, the consistency of two patterns verifies the correctness of the derivation from Eq.\ref{a4} to Eq.\ref{a7}.

\begin{table}[htb]
\setlength{\belowcaptionskip}{0.3cm}
\centering
\caption{parameter list of the PAT antenna}
\label{t1}       
\begin{tabular}{l|llll}
\hline\noalign{\smallskip}
Parameter &$f$&$l_{0}$&$a$ &$\beta$ \\
\noalign{\smallskip}\hline\noalign{\smallskip}
Value &$60 GHz$&$40$&$80 mm$&$\frac{2\pi}{9}$\\
\noalign{\smallskip}\hline
\end{tabular}
\end{table}

\begin{figure}[htb]
 \centering
 \includegraphics[width=3.5in]{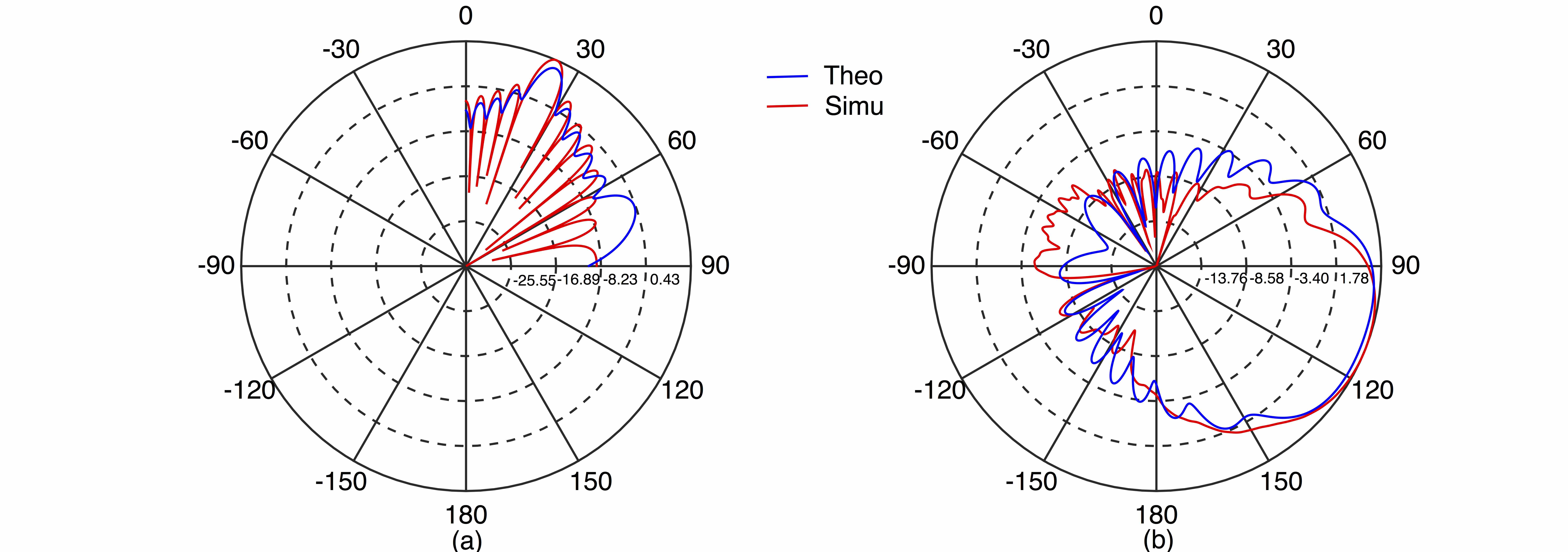}
 \caption{Simulated and theoretical results of relative amplitude: (a) results in zenith direction (b) results in azimuth direction}
 \label{p3}
\end{figure}

The antenna is simulated in the commercial software CST. The maximum point of the farfield is supposed at point $P(\theta_{s},\varphi_{s})$. Fig.\ref{p3}(a) shows the simulated radiation pattern when $\varphi=\varphi_{s}$ and the theoretical one calculated by Eq.\ref{a7} when $\varphi=\varphi_{0}$. The simulated and theoretical radiation patterns agree well, and the main lobe direction, which are $24.0^{o}$ and $23.8^{o}$ respectively, are also well matched with an error less than $0.2^{o}$.
Fig.\ref{p3}(b) shows the simulated and theoretical result of relative amplitude distribution in azimuthal direction when $\theta=\theta_{s}$. The patterns are consistent on the whole and the simulated result of $\varphi_{s}$ is $112.0^{o}$. There is a $2.0^{o}$ deviation compared with the theoretical value $\varphi_{0}=\frac{\beta}{2}+\frac{\pi}{2}=110.0^{o}$.

\begin{figure}[htb]
 \centering
 \includegraphics[width=2.5in]{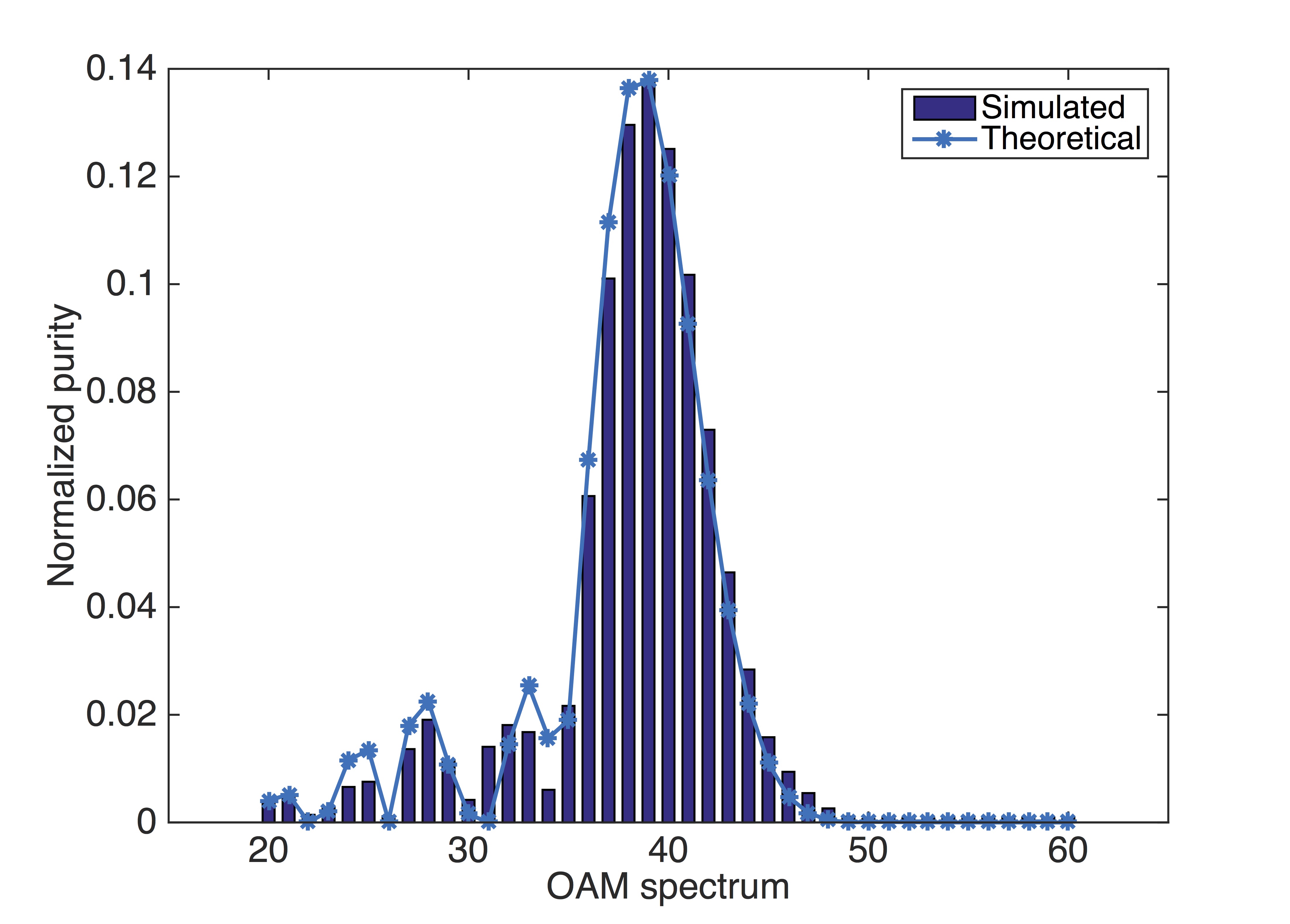}
 \caption{Theoretical and simulated results of OAM spectrum of this PAT antenna}
 \label{p4}
\end{figure}

The OAM spectrum in the main lobe direction in the $\theta$ direction is of great significance in the applications of the MG generated by PAT antenna. The theoretical OAM spectrum and the calculated one by the simulated field distributions are shown in Fig.\ref{p4}. 

Clearly, the theoretical and simulated results are well matched either. Simulated OAM spectrum can be identified as the integer sampling of the theoretical OAM spectrum function. The acceptable error between the theoretical and simulated results comes from the imperfect antenna design.
\section{Arbitrary non-uniform OAM source scheme}

\begin{figure}[htb]
 \centering
 \includegraphics[width=3.0in]{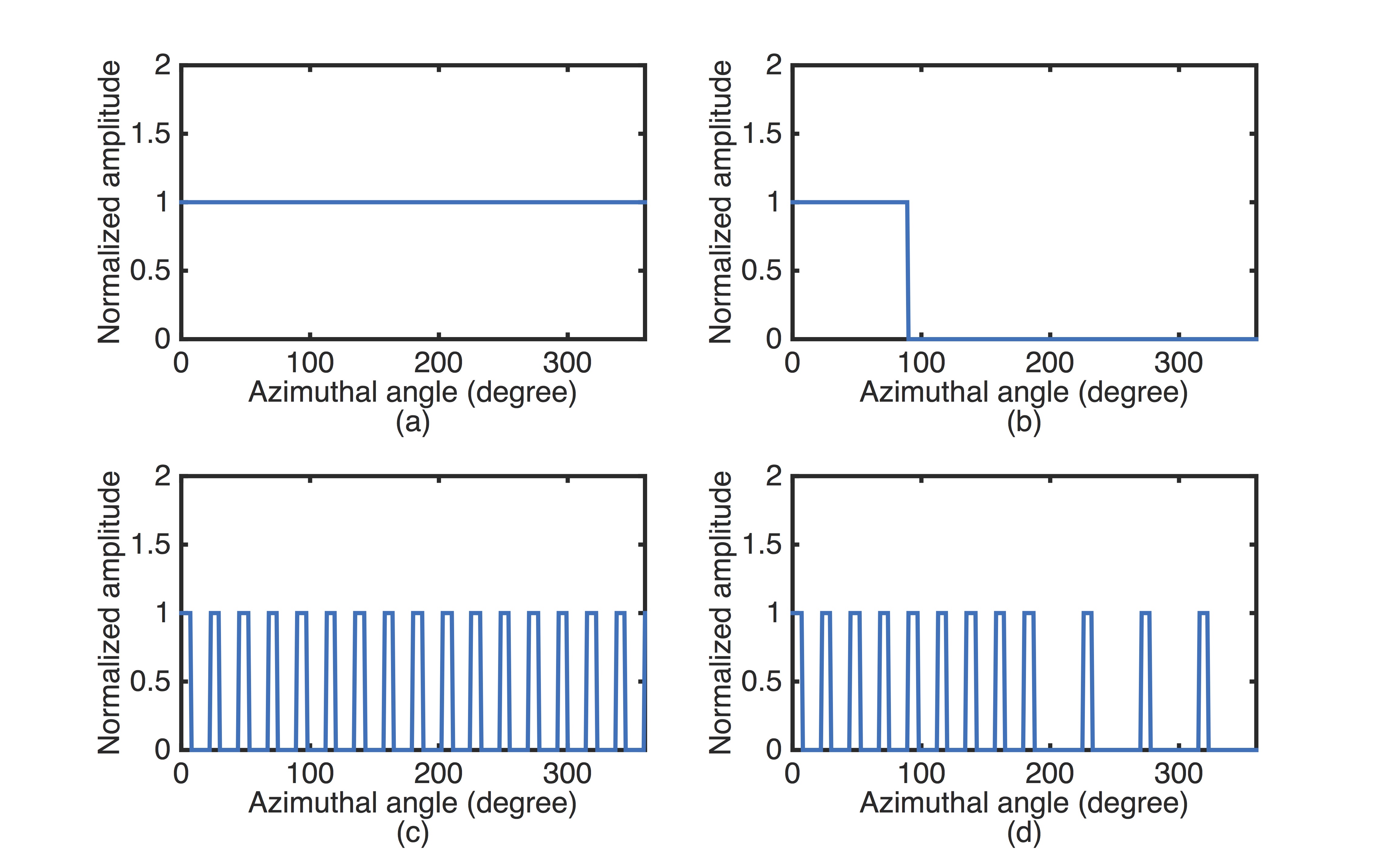}
 \caption{The amplitude distributions of some classical OAM sources: (a) uniform OAM source (b) the PAT source (c) the UCA source (d) the DWCA source}
 \label{p5}
\end{figure}

For arbitrary non-uniform OAM source who satisfies $\vec I_{n}=I_{n}(\varphi')e^{-jl\varphi'}$, $I_{n}(\varphi')$ is the amplitude distribution function according to the antenna. Fig.\ref{p5} shows the amplitude distributions of some classical OAM sources, including the uniform OAM source, the PAT source, the uniform circular arrray (UCA) source and the density-weighted circular array (DWCA) source in \cite{ISI:000492854000062}. 

The radiation field can be given by:
\begin{equation}
\label{a10}
\begin{split}
 \vec E_{n}(\theta,\varphi)=\sum_{l=-\infty}^{\infty} M\lbrack I_{n}(l)\rbrack J_{l}(kasin\theta)e^{-jl(\varphi-\frac{l}{|l|}\frac{\pi}{2})}
 \end{split}
\end{equation}
where $M\lbrack I_{n}(l)\rbrack$ is the FT of the amplitude distribution. From the electric field expression, the characteristics of the beam, such as the main lobe direction $\theta_{m}$ in the $\theta$ direction, $\varphi_{m}$ in the $\varphi$ direction and the OAM spectrum $A_{m}$ at any plane vertical to the transmission axis can be calculated. 

\section{Conclusions}
In this paper, a non-uniform OAM source is demonstrated to generate the OAM MG. Its analytical solution of the far field radiation is obtained by the FT of the source. The non-uniform OAM source can be equivalent to a sum of a series of uniform OAM source with amplitude modulation, whose weights are exactly the FT of the amplitude distribution of the non-uniform OAM source. Taking the PAT scheme as an example, the mainlobe and the OAM spectrum are calculated from the radiation field expression and compared with a simulated antenna. The calculated result shows good consistency with the simulated one, which verified that the proposed method can provide a theoretical guidance for the design of the OAM MG antenna. 

\vskip3pt
\ack{The authors acknowledge the funding supported from the National Natural Science Foundation of China under Grant number 61571391.}

\vskip5pt

\noindent Shilie Zheng (\textit{Zhejiang University, Hangzhou, China})
\vskip3pt

\noindent E-mail: zhengsl@zju.edu.cn
\bibliographystyle{iet}
\bibliography{ref}

\begin{thebibliography}{1}

\bibitem{Mohammadi2010Orbital}
Mohammadi, S.M., Daldorff, L.K.S., Bergman, J.E.S., Karlsson, R.L., Carozzi,
  T.D.: `Orbital angular momentum in radio—a system study', \emph{IEEE
  Transactions on Antennas and Propagation},  2010, , (2), pp.~565--572

\bibitem{Zheng2018Realization}
Zheng, S., Chen, Y., Zhang, Z., Jin, X., Chi, H., Zhang, X., et~al.:
  `Realization of beam steering based on plane spiral orbital angular momentum
  wave', \emph{IEEE Transactions on Antennas and Propagation},  2018,

\bibitem{cl2020MGmimo}
Xiong, X., Zheng, S., Zhu, Z., Yu, X., Jin, X., Zhang, X.
\newblock `Performance analysis of plane spiral oam mode-group based mimo
  system'.
\newblock In: IEEE Communications Letters. (IEEE,  accepted.

\bibitem{Jack2008Angular}
Jack, B., Padgett, M.J., Franke.Arnold, S.: `Angular diffraction', \emph{New
  Journal of Physics},  2008, \textbf{10}, (10), pp.~6456--6460

\bibitem{xiong2020oam}
Xiong, X., Zheng, S., Zhu, Z., Chen, Y., Wang, Z., Yu, X., et~al.. `Oam
  mode-group generation method: Partial arc transmitting scheme'. (,  2020

\bibitem{ISI:000492854000062}
Liu, K., Cheng, Y., Wang, H., Yang, Q.
\newblock `{An OAM-generating Method Using Density-weighted Circular Array}'.
\newblock In: {Knott, P}, editor. {2019 20TH INTERNATIONAL RADAR SYMPOSIUM
  (IRS)}. ({IEEE},  {2019}.

\bibitem{JhaFourier}
Jha, A.K., Jack, B., Yao, E., Leach, J., Boyd, R.W., Buller, G.S., et~al.:
  `Fourier relationship between the angle and angular momentum of entangled
  photons', \emph{Physical Review A}, , \textbf{78}, (4), pp.~043810

\end{thebibliography}
\end{document}